\documentclass[nofootinbib,onecolumn,secnumarabic,showpacs,preprintnumbers,amsmath,amssymb,nobibnotes,12pt,pra]{revtex4-1}
\usepackage{color}
\usepackage{graphicx}
\newcommand\go{$g^{(2)}(0)\,\,$}
\newcommand\gt{$g^{(2)}(\tau)\,\,$}
\begin{document}
\title{
Time-delayed intensity-interferometry of the  emission from ultracold atoms in a steady-state magneto-optical trap}
\author{Muhammed Shafi K.$^{1}$, Deepak Pandey$^{1,2}$,  Buti Suryabrahmam$^{1}$, B.S. Girish$^{1}$, and    
Hema Ramachandran $^{1,*}$}
% etc
\address{$^{1}$ Raman Research Institute, C.V. Raman Avenue, Sadashivnagar, Bangalore, INDIA-560080\\
 $^2$ Presently at LP2N, Universite Bordeaux, France}

%\email[$^*$]{hema@rri.res.in}}

\date{\today}
\begin{abstract}
An accurate measurement of the bunching of photons in the fluorescent 
emission from an ultracold ensemble of thermal $^{87}Rb$ atoms in a 
steady-state magneto-optical trap is presented. Time-delayed-intensity-interferometry (TDII) 
performed with a  5-nanosecond time resolution yielded a second-order intensity correlation function that 
has the ideal value of 2 at zero delay, and that shows coherent Rabi oscillations 
of upto 5 full periods -  much longer than the spontaneous emission lifetime of the 
excited state of Rb.  The
oscillations are damped out by $\sim150$ns, and thereafter, as expected from a thermal source, an
exponential decay is observed, enabling the determination of the     
temperature of the atomic ensemble. Values  so obtained compare well with those determined 
by standard techniques. TDII
thus enables a quantitative  study of the coherent and 
incoherent dynamics, even of a large thermal ensemble of atomic emitters.
   
\end{abstract}

\pacs{ } % end of PACS codes

\maketitle

%\section{Introduction}

\begin{figure*}
\includegraphics[height = 10.5cm, width =15cm]{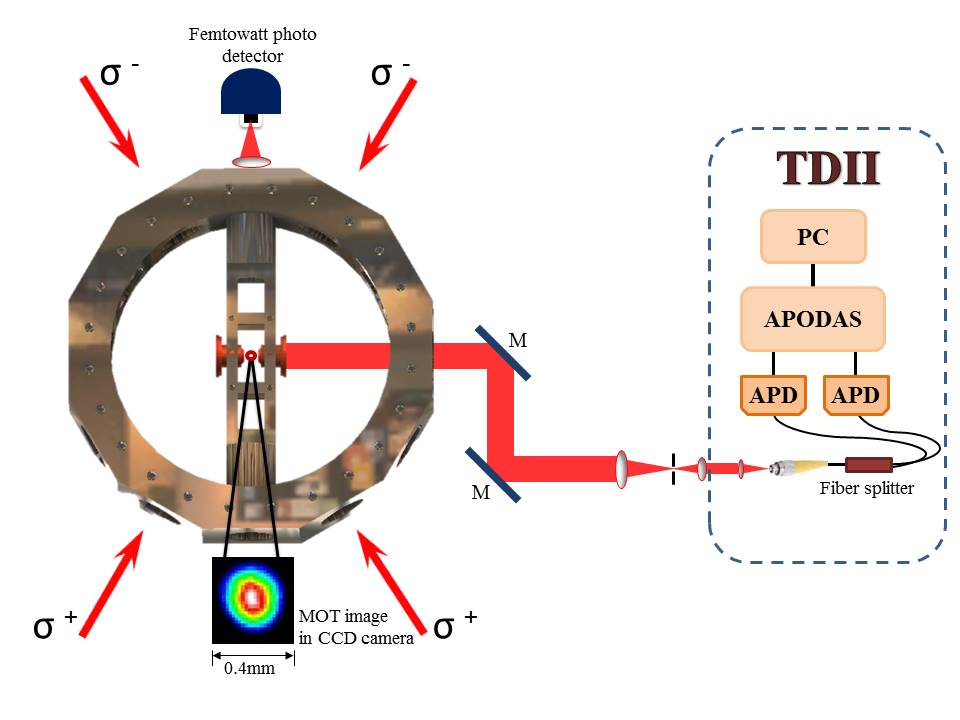}
\caption{\it Schematic of the experiment showing the collection of 
cold atoms inside the MOT  chamber. A lens within the chamber collects 
fluorescence from the atoms, which is passed onto the TDII setup where the 
arrival of individual photons are time-stamped and archived. Bold arrows show the cooling beams in the x-y plane. The z-beams (perpendicular to the page) are 
not shown.  }
\label{Schematic}       
\end{figure*}

Intensity interferometry,  or  the measurement of  photon correlation,  provides  a wealth of 
information regarding the light source and the mechanism of emission. A study of the 
mere arrival 
times of photons from the  source enables one to 
determine its nature - whether it is  thermal (chaotic),  or  coherent, or  quantum.
 The form of approach to the 
asymptotic  value  allows one to infer further details of the source. For example, 
one of the earliest measurements \cite{HBT} of spatial photon correlation of 
light from the star Sirius enabled the determination of the diameter of the star. 
Intensity interferometry is routinely used in particle physics \cite{Baym, particle-physics} to study 
decay processes  and to deduce the interaction between particles. 
Other applications include the search for naturally occuring non-classical sources of radiation in 
astrophysics \cite{stellar},  study of light emission from nanostructures 
\cite{Bayer} 
and particle size measurements \cite{DLS}. 
Typically, pairs of detectors either seek the simultaneous arrival of photons,  or 
measure the delayed arrival of a photon at one detector 
with respect to the arrival of a photon at the other.   The measurements are 
quantified by  the  second order correlation function, (also known 
as intensity-intensity correlation function),  $g^{(2)}(R, \tau)  = \frac{\Big  {\langle I_1(r, t)\Big \rangle \Big\langle I_2(r+ R, t + \tau) \Big \rangle}}
{\Big{\langle I_1(r,t) \Big \rangle \Big \langle I_2(r, t)\Big \rangle }}$,  where $I_1(r, t)$ and 
$I_2(r + R, t+\tau)$ are the 
intensities of light reaching detectors D1 and D2 at locations r and r+R at times 
$t$ and $t + \tau$  and 
the angular brackets denote time averaging. It is the correlation of the number of 
photons, or intensities, that is examined,  as opposed to a correlation of 
amplitudes in conventional interferometers, and hence the  name time-delayed 
intensity-interferometry (TDII).  It has been theoretically shown that (see, for example \cite{Baym}) that the second order correlation function $g^{(2)}(\tau 
= 0)$ =  2 for a thermal state, implying a tendency for bunched or
correlated emission of photons; is unity for a coherent source, implying emission 
of 
photons at random times, and equals  $1-\frac{1}{n}$ for a n-photon Fock state, 
signifying 
anti-bunching. The value of \gt for  all sources, however, approaches 
unity for long time delays.  \\ 
\indent Temporal bunching of photons from thermal sources, ever since the postulation of the concept, has been an intriguing phenomenon, and has been the focus 
of  numerous experiments. The earliest  laboratory thermal source 
studied was a Hg vapour lamp \cite{Hg1}, light from  which showed a meagre bunching 
of 1.17. Martiensen and Spiller  devised a method of creating pseudo-thermal light   by 
transmitting coherent laser light through a rotating  ground glass 
plate \cite{GroundGlass} such that the time-varying surface inhomogenieties  introduced 
temporal and spatial 
decoherence. In recent years, 
laboratory 
control and measurement 
techniques have enabled creation of pseudo-thermal light sources with 
theoretically expected values of \gt = 2   \cite{Anders, AB-EPL, JOSA,  EPJP}.  \\
\indent We report here  Time-Delayed  Intensity-Interferometric measurements 
on light from another source of bunched photons - an ensemble of  laser cooled 
atoms. Though laser cooled atoms have been 
available for more than  three decades,  direct TDII measurements of 
their emission  have been very few\cite{Aspect, Bali, Nakayama}.  
All measurements hitherto have been carried out in  optical molasses, which were 
either periodically, or continuously loaded 
with atoms precooled in a MOT.   In this paper we present measurements of 
the second order correlation function of light emitted by ultracold atoms in a 
steady-state magneto-optical trap, where the cooling and repumper beams, and also 
the quadrupolar magnetic field are kept on. We observe the ideal value of 2 for the zero-delay intensity-intensity correlation function.  Damped Rabi oscillations are observed for time delays upto 
$\sim 150$ns, and an exponential decay for longer time delays. 
Despite the fact that the  emission being studied is from a 
collection of uncorrelated atoms that are in random thermal motion, and that the 
observations are averaged over an 8-hour period, coherent effects are seen, 
bringing out the power of  higher order correlations in revealing hidden 
periodicities and providing a measure of coherent and incoherent dynamics.   The 
exponential decay at long time delays was used to determine the temperature of 
the ensemble.   \\
\indent Time delayed Intensity Interferometry was performed on the fluorescent emission from  $^{87}$Rb atoms cooled and trapped in a  
magneto-optical trap (MOT) (Fig. \ref{Schematic}), which differed from 
usual MOTs, in that the two pairs of beams in the x-y plane were steeply inclined 
to each other, enclosing  an angle of 55$^\circ$ rather than the usual 90$^\circ$ so 
as to accomodate, within the chamber,   a pair of lenses of short-focal length and 
high
numerical aperture. These  lenses were  positioned facing each other, such that 
their focal points coincided with the centre of the MOT and could thus be used 
to focus light onto the MOT, or collect light emitted by a small volume within 
the cold cloud.  To avoid clipping at the lens mounts leading to undesired 
scattered light, the diameters of the beams in the x-y plane were restricted to 
$\sim$1.5mm while the z-beam had a diameter of 8mm. 
The  cooled and trapped atoms were viewed using a CCD camera, 
and their number  estimated by collecting part of the fluorescent light
onto a femtowatt detector.  The typical cloud was roughly ellipsoidal 
with a mean diameter of $\sim400\mu$m, and contained about 20000 atoms.  

 An intensity-interferometer was  formed using a  fiber splitter, where the two 
output ends were connected to two high speed   APDs  that had  
quantum efficiencies  of 65$\%$ and a dead 
time of 30ns. The output of the APDs were fed to a homebuilt, FPGA-based time-tagged-single-photon recorder, APODAS (Avalanche Photodiode Optical 
Data Acquisition System  \cite{arXiv} that utilised high speed ethernet 
connectivity and  stored, in a PC in realtime, the arrival times  of all detected photons 
with a temporal  resolution of 5 nanoseconds.  Post-processing
by software enabled the determination of \gt  for all $\tau$ from a
single recording of the data \cite{EPJP}. 
As we worked in the photon-counting mode, the expression for \gt  in terms of coincidences is \cite{Bali}
\begin{equation}
g^{(2)}(\tau) = N_{c} T/ (N_1 N_2  \tau_c)
\end{equation} 
where $\tau$ at is the delay between arrival at the two detectors, $N_1 , N_2$ and 
$N_c$ are the number of counts at 
detector 1, detector 2
and the coincident counts respectively. T is the total observation time and 
 $\tau_c$  the  time window for coincidence (arrival of  two photons 
is considered simultaneous if they are
detected within a time gap of $\tau_c$). 

\begin{figure*}
\vspace{-2cm}
\includegraphics[height = 12cm, width =12cm]{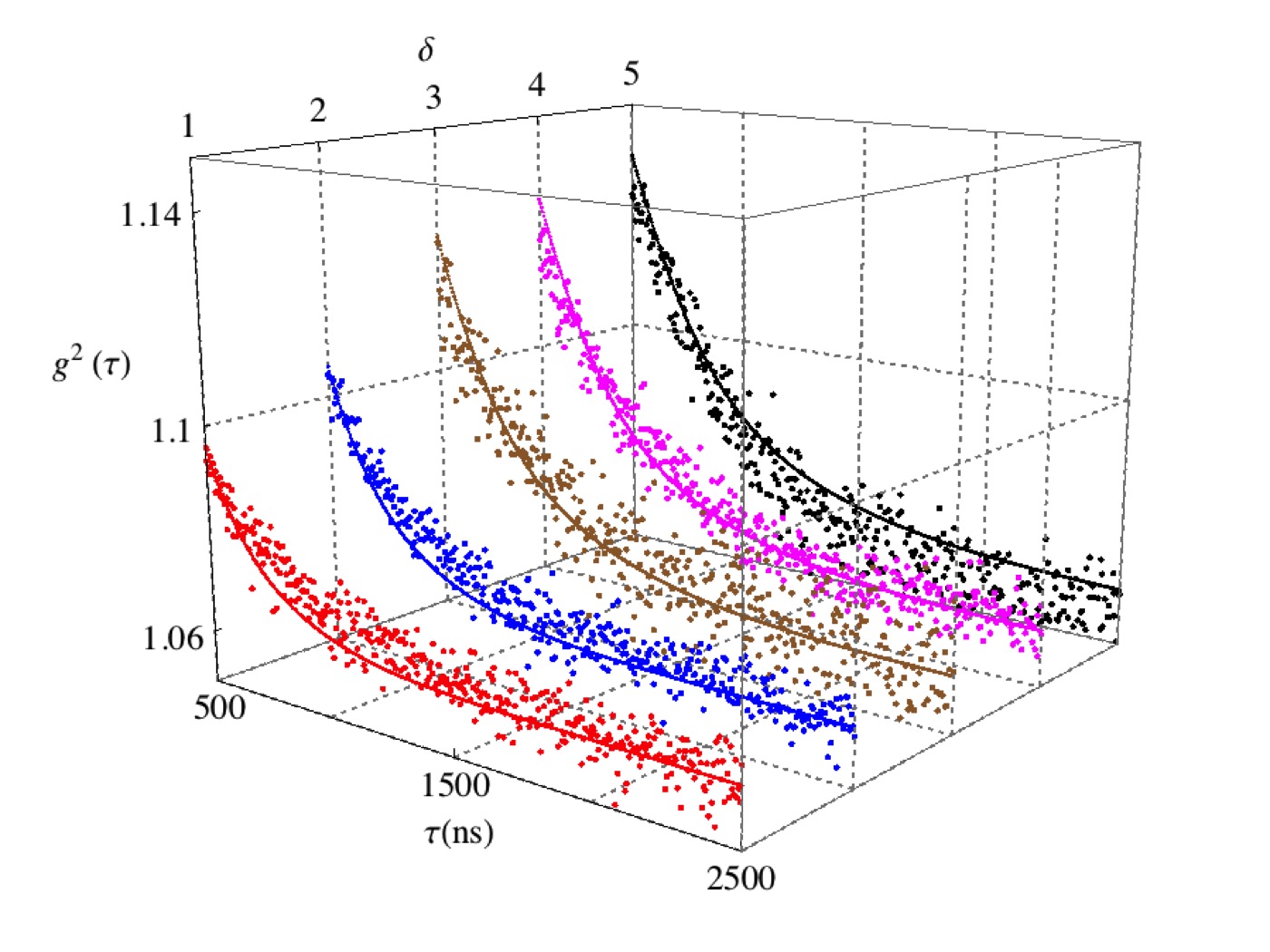}
\vspace{-1cm}
\caption {\it{\gt vs $\tau$  for different  detunings, $\delta$  
(1 = -14.5MHz, 2 = -16.1MHz, 3 = -18MHz, 4 = -19.4MHz, 5 = 
-22MHz) of the cooling beam from the cooling transition. The values obtained from TDII measurements are shown as dots. The 
solid curves represent the values of \gt  obtained from Eqs.\ref{Doppler} and 
\ref{eq-g2-g1-D}, with temperatures $100\mu$K, $80\mu$K, $60\mu$K, $50\mu$K, $43\mu$K, for the 5 values of detunings,  (1 = -14.5MHz, 2 = -16.1MHz, 3 = -18MHz, 4 = -19.4MHz, 5 = 
-22MHz) respectively. The data for 
all sets are fit with the same choice of parameters, to obtain the temperature.}}
\label{temp}       
\end{figure*}

\begin{figure*}
\includegraphics[height = 8cm, width =10cm]{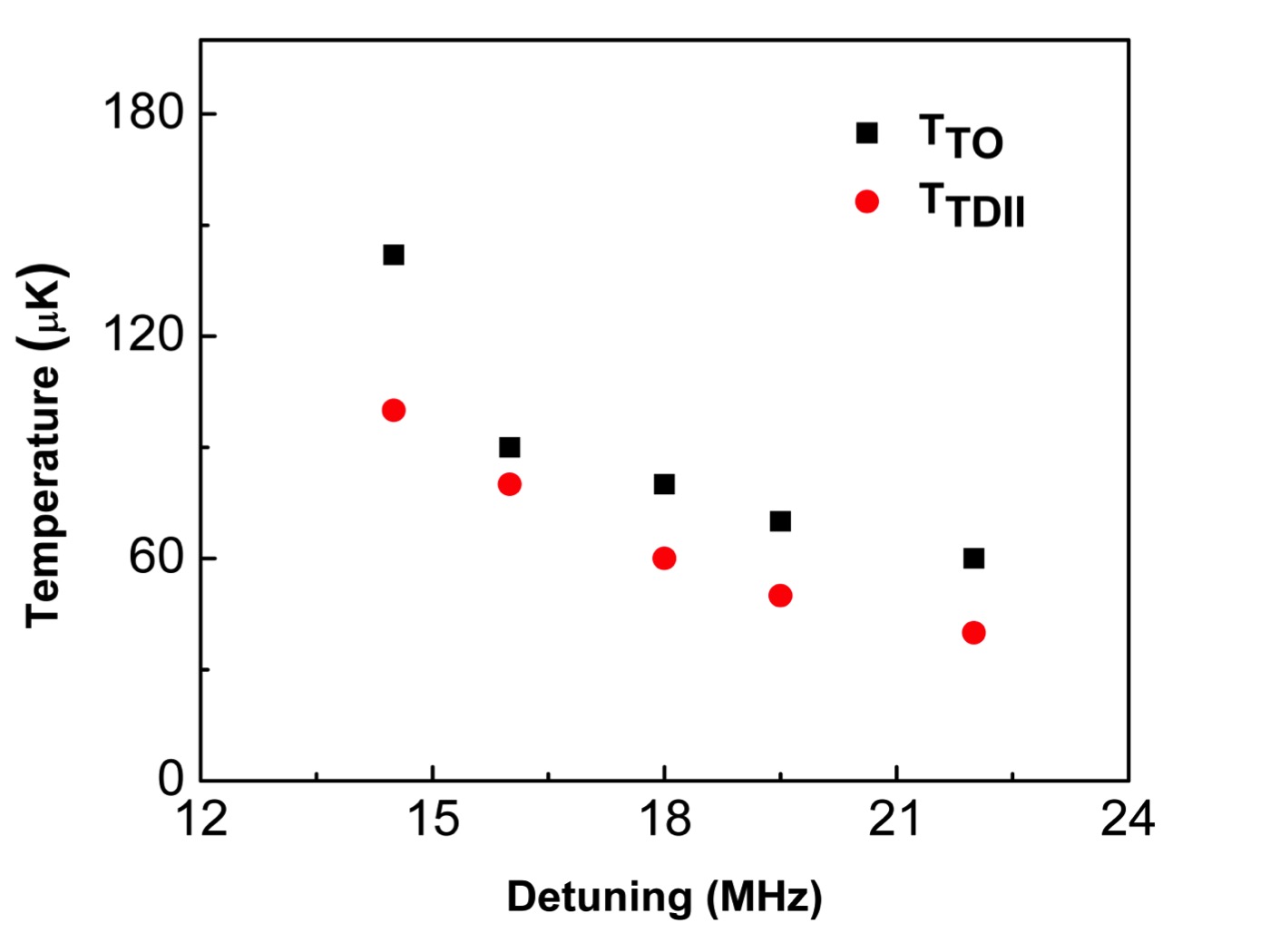}
\caption{\it Comparison of the temperature of the cold atoms, 
 obtained from two methods - TDII (circles)  and 
trap oscillation  (squares) -  as function of the (red) detuning of the cooling beam.}
\label{temp-compare}       
\end{figure*}

       For determining the bunching characteristics of emission from an
ultracold atomic ensemble, $^{87}$Rb atoms were laser cooled  from close to 
room temperature, to $\sim 100 \mu$K. 
The MOT was extremely stable, with the lasers locked 
and the cold cloud in steady-state for  days. A typical run of the experiment 
lasted 8 hours. The cold cloud was obtained, and the cooling and  repumper 
beams, and the magnetic field were kept
switched on for the entire duration of the experiment.  The cold ensemble was 
constantly monitored by imaging it on a camera, and also by measuring the 
fluorescence on a femtowatt detector (see Fig. \ref{Schematic}). For the purpose 
of determining 
its second order correlation function, fluorescence light 
from the central region of the cloud was collected by the high-numerical 
aperture lens
placed inside the sample chamber, a few millimeters from the trap centre. Care 
was taken to ensure that no part of the laser light
entered this lens, either directly, or upon being scattered by the parts of the 
MOT chamber.  The light was conveyed by  a series of mirrors 
and lenses to the input of the TDII setup. The count registered in the presence of the cold cloud was in 
the 
range 40,000 - 80,000/s  while in the absence
of the cold cloud it reduced to $\sim$1200/s,  confirming that it is 
predominantly light  from the cold atoms that
enters the TDII setup. Photon arrival time data was recorded for 
various detunings of the cooling beam. For each detuning the number 
of 
atoms trapped was estimated from the fluorescent intensity recorded 
on 
the femtowatt detector. While the total number of atoms collected ranged 
from $\sim8000$ to $\sim22000$, the high numerical 
aperture lens employed for accepting light for TDII measurement
restricted  the collection of light to that 
from approximately one-tenth of the volume of the cold ensemble.
The temperature of the collection of atoms 
was determined by  the trap oscillation method \cite{trap-osc}, and
 was found to  range from $200\mu$K to  
$50\mu$K  for 
detunings of the cooling laser varying from  -12MHz to -22MHz. The 
effective Rabi frequencies ranged from 25MHz to 40MHz.  \\
\begin{figure}
\vspace{-2cm}
\includegraphics[height=12cm, width=16
cm]{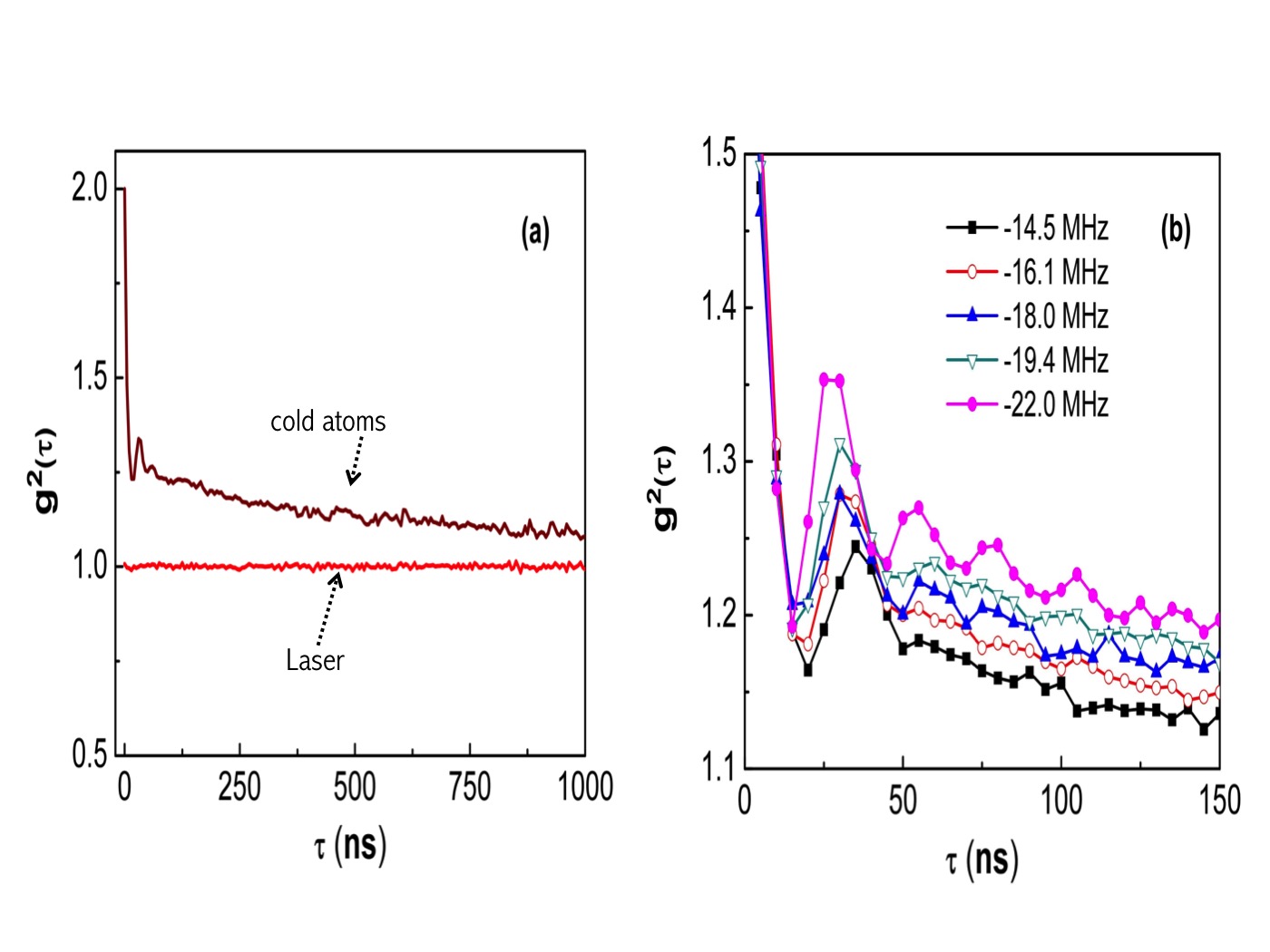}
\caption{{\it (a)  Experimentally obtained $g^{(2)}(\tau)$ vs $ \tau$  for  laser light measured before its 
entry to  the MOT chamber (red) and  for fluorescence light from the cold atoms (brown), and
(b) Oscillations seen  \gt at short time scales, for different detunings of the cooling beams; the curves are drawn as a guide to the eye, and the values  of the detuning are given in the legend.}}
\label{g2}       
\end{figure}
\indent Two-time intensity-correlation values were derived from  the TDII measurements. Fig. 
\ref{g2}a shows the second order correlation functions of the input laser light 
determined before it enters the MOT chamber, and of the light emitted by the 
ultracold atoms trapped in the MOT. It is clear that  absorption and re-emission by the 
thermal ensemble of atoms has led to bunching in light, which was initally 
coherent (\gt = 1 for all $\tau$). 
Several interesting features are observed in \gt for light from 
the cold atoms. Periodic oscillations, are seen; these are damped out by 
$\sim 150$ns, and thereafter the curve decays steadily towards unit value. The 
curve \gt is less noisy for small $\tau$ than for larger ones. 
 As 
the cooling beams are further detuned from the $5S_{1/2}F=2 \rightarrow 
5P_{3/2}F=3$ transition (the so-called "cooling transition"),  
the value of \go approaches 
closer to 2   while the oscillations in correlation  are more prominent and more rapid, 
and their damping is slower (Fig. \ref{g2}b). \\
\indent Let us, for simplicity, assume the Rb 
atom to be a 2-
level system. Irradiation of an atom by coherent, near resonant light  causes two 
processes to occur. One is the periodic absorption and  coherent emission of 
radiation leading to an oscillation between the excited and the ground state at the  
Rabi frequency  - a rate determined by the intensity and detuning of the incoming 
radiation. The other is the absorption and spontaneous random 
emission, at a  rate that falls exponentially with a characteristic lifetime, which, for  
the excited state of  $^{87}$Rb  is $\sim$28ns. 
The atoms in the ensemble, being uncorrelated, do not emit in unison, and thus 
no periodicity will be evident in the direct observation of emission from the 
collection. However, all atoms undergo Rabi oscillations at the same 
frequency, and thus have a high probability of emission at the same regular 
interval. The second order correlation, which is a measure of the probability of 
emission at time $t + \tau$ conditioned on an emission having occurred at time $t$,
will therefore exhibit  periodic maxima at regular intervals $\tau_R$,  the inverse of the Rabi frequency. Thus, the two-time  intensity 
correlation measurement is a  simple yet powerful technique that 
can reveal hidden periodicities. The periodic oscillations arising from 
coherent dynamics under steady-state driving  fields, nevertheless, show decay as  decoherence 
sets in due to spontaneous emission and 
inter-atomic collisions. 
This reasoning  also explains why 
\gt  is less noisy  for short time delays than for larger delays. The 
Rabi frequency in our experiment being a few tens of MHz, coherent oscillations 
should occur at the time scales of  few tens of nanoseconds. The lifetime of the 
excited state is $\sim 28$ns, implying that a decay 
in amplitude of oscillation will occur over $\sim100$ns ( a few lifetimes). 
Interatomic collisions occur at yet longer time scales. Thus, short delays show
cleaner curves. \\
Let us consider a thermal collection of $N$ independent atoms 
under the action of a near-resonant  driving field of Rabi frequency $\Omega$.  
The 
electric field $ {\mathcal{E}}(\vec{r})$ due to coherent emission, at a 
point of observation  $\vec{r}$ is the resultant of contributions from each atom, $j$, and may be written as (Eq. 13.48 in \cite{GSA})
\begin{equation}
  {\mathcal{E }}(\vec{r}) \, \,\,\,= \, \,\,\,\displaystyle\sum_{j =1}^{N}  {\mathcal{E}_j}  \, \,\,\, \sim \,\,\,\, \displaystyle\sum_{j =1}^{N} K_j {S_j }(e \rightarrow g)e^{i \phi_j} 
\label{eq-emission}
\end{equation}
 where $S_j(e \rightarrow g)$,  is the de-excitation operator for the two-level atom,  the dynamics of which is given by the master equation for the driven two-level atom (Eq. 13.1 in \cite{GSA}), and 
$\phi_j$  the phase of the  
 electric vector  (due to the coherent emission of the $j^{th}$ atom) at the point of observation (detection) depends on the location of the atom and the orientation
of the atomic dipole. 
In a MOT, the resultant driving field due to the six cooling beams varies in a complex manner in 
intensity and polarization  from position to position. Likewise, the orientation of each atom 
varies as it moves within the MOT region (see for example \cite{Gomer}). \\
  The second order intensity correlation function is 
defined as :
\begin{equation}
  g^{(2)}(\tau) = \frac{\langle\mathcal{E}^\dagger(0)\mathcal{E}^\dagger(\tau)\mathcal{E}(\tau)\mathcal{E}(0) \rangle} {\langle\mathcal{E}^\dagger(0)\mathcal{E}(0)\rangle^2}
\label{eq-g2}
\end{equation}
Substituting from Eq.\ref{eq-emission} to Eq.\ref{eq-g2} yields terms of the following forms, with appropriate prefactors : 
\begin{align*}
&(a) \hspace{0.5cm}\displaystyle\sum_{j=1}^{N}\langle \mathcal{E}_j^\dagger(0)\mathcal{E}_j^\dagger(\tau)\mathcal{E}_j(\tau)\mathcal{E}_j(0)\rangle\\
&(b) \hspace{0.5cm}\displaystyle\sum_{i=1}^{N}\displaystyle\sum_{j=1, j\neq i}^{N}\langle \mathcal{E}_i^\dagger(0)\mathcal{E}_j^\dagger(\tau)\mathcal{E}_j(\tau)\mathcal{E}_i(0)\rangle\\  
&(c) \hspace{0.5cm}\displaystyle\sum_{i=1}^{N}\displaystyle\sum_{j=1, j\neq i}^{N}\langle \mathcal{E}_i^\dagger(0)\mathcal{E}_j^\dagger(\tau)\mathcal{E}_i(\tau)\mathcal{E}_j(0)\rangle\\  
&(d) \hspace{0.5cm}\displaystyle\sum_{i=1}^{N}\displaystyle\sum_{j=1, j\neq i}^{N}\langle \mathcal{E}_i^\dagger(0)\mathcal{E}_i^\dagger(\tau)\mathcal{E}_j(\tau)\mathcal{E}_j(0)\rangle\\  
&(e) \hspace{0.5cm}\displaystyle\sum_{i=1}^{N}\displaystyle\sum_{j=1, j \neq i}^{N}\displaystyle\sum_{k=1, k\neq i, j} 
 ^{N}\langle\mathcal{E}_i^\dagger(0)\mathcal{E}_j^\dagger(\tau)\mathcal{E}_k(\tau)\mathcal{E}_i(0)\rangle \text{\,\,and various permutations }\\ 
&(f) \hspace{0.5cm}\displaystyle\sum_{i=1}^{N}\displaystyle\sum_{j=1, j \neq i}^{N}\displaystyle\sum_{k=1, k\neq i, j} ^{N}\displaystyle\sum_{l=1, l\neq i,j,k}^{N}\langle\mathcal{E}_i^\dagger(0)\mathcal{E}_j^\dagger(\tau)\mathcal{E}_k(\tau)\mathcal{E}_l(0)\rangle
\end{align*}
Terms of the form  (a) represent single-atom contributions. At $\tau 
=0$ these show  antibunching - an atom cannot emit more than one photon at a time. \\
Recognising that the emitters are a thermal collection of uncorrelated atoms, the 
operators in the terms in (b) may be re-ordered and factorised to yield : 
\begin{align*}
(b) &\corresponds  \displaystyle\sum _{i=1}^{N} \displaystyle\sum _{j=1, j \neq i}^{N} [ \langle\mathcal{E}_i^\dagger(0)\mathcal{E}_i(0)\rangle
\langle\mathcal{E}_j^\dagger(\tau)\mathcal{E}_j(\tau)\rangle]\\
&=  N (N-1)  \frac{I^2}{N^2}
\end{align*}
where $I$ is the total intensity at the detector  due to $N$ atoms.  On similar lines, terms in (c) lead to the auto-correlation :
\begin{align*}
(c)  &\corresponds    N(N-1) 
|\langle\mathcal{E}_j^\dagger(\tau)\mathcal{E}_j(0)\rangle|^2
\end{align*}
Terms consituting (d) are related  to the anomalous correlation, which, for a thermal cloud, vanish on time averaging. Similarly, terms (e) and (f), 
due to the random phases, also drop out upon time averaging. From these arguments, we now find that 
\begin{align*} 
\langle\mathcal{E}^\dagger(0)\mathcal{E}^\dagger(\tau)\mathcal{E}(\tau)\mathcal{E}(0) \rangle & = N [ \langle \mathcal{E}_j^\dagger(0)\mathcal{E}_j^\dagger(\tau)\mathcal{E}_j(\tau)\mathcal{E}_j(0)\rangle] 
\, \, \, + \,\,\,N(N-1)\Big{[}\frac{I^2}{N^2} \,\, +\,\, \frac{I^2}{N^2} \Big{|} \frac{N\langle \mathcal{E}_j^\dagger(\tau)\mathcal{E}_j(0)\rangle}{I}
\Big{|}^2 \Big{]}
%& =  \,\,  \frac{N^2 \mathcal{I}^2[ \langle \mathcal{E}_j^\dagger(0)\mathcal{E}_j^\dagger(\tau)\mathcal{E}_j(\tau)\mathcal{E}_j(0)\rangle]}{N\mathcal{I}^2} + \langle N \rangle ^2  \mathcal{I}^2 \Big{[} 1 + \Big{|} \frac{\langle \mathcal{E}^\dagger(\tau) \mathcal{E}(0)\rangle}{\mathcal{I}}\Big{|}^2\Big{]} 
\end{align*}
Denoting by  $\mathcal{I} = I/N$, is the intensity due a single atom, one obtains 
\begin{equation}
g^{(2)}(\tau) \,\, = \,\,    \frac{\langle \mathcal{E}_j^\dagger(0)\mathcal{E}_j^\dagger(\tau)\mathcal{E}_j(\tau)\mathcal{E}_j(0)\rangle}{N\mathcal{I}^2} + \Big{(}1 - \frac{1}{N}\Big{)}\Big{[}1 + \Big{|} \frac{\langle \mathcal{E}^\dagger(\tau) \mathcal{E}(0)\rangle}{\mathcal{I}}\Big{|}^2 \Big{]}
\end{equation}
The first term represents the single-atom contributions, which diminishes when the 
number of atoms becomes large. In this limit, the above equation   leads to 
the well known relation between the first and second order correlations : 
\begin{equation}
g^{(2)}(\tau) \,\, = \,\,   1 + \Big{|} g^{(1)}(\tau)\Big{|}^2 
\label{eq-g2-g1}
\end{equation}
As is well known from the Wiener-Khinchtine theorem, the fourier transform of 
$g^{(1)}(\tau)$ yields the power spectral density of the emission. In the time 
domain it leads to Rabi oscillations.
\begin{equation}
g^{(1)}(\tau) =  A_0 \,\,+  \,\,
A_1e^{- \Gamma_0 \tau}  \,\,+  \,\, A_2e^{i \Omega \tau - \Gamma_1 \tau}  \,\,+ \,\, A_3e^{-i \Omega \tau -\Gamma_2 \tau} 
% \text{\,\,+ incoherent terms\\ + high frequency terms}
\label{final}
\end{equation}
Here $\Omega = (\Omega _0^2 + \delta^2)^{1/2} $, is the effective Rabi frequency at detuning $\delta$. 
The second order correlation function, given by Eq. \ref{eq-g2-g1} would then have the form \\
\begin{eqnarray*}
 g^{(2)}(\tau) =  A_0 \,\,+  \,\,
A_1e^{- \Gamma_0 \tau}  \,\,+  \,\, A_2e^{i \Omega \tau - \Gamma_1 \tau}  \,\,+ \,\, A_3e^{-i \Omega \tau -\Gamma_2 \tau}\\
 + \,\,A_4e^{2i \Omega \tau - \Gamma_3 \tau} 
   + \,\, A_5e^{-2i \Omega \tau -\Gamma_4 \tau } \text{ + incoherent terms}
\label{final2}
\end{eqnarray*}
Thus, the emission from a  collection of N atoms
will display  coherent Rabi oscillations that decay at rates 
indicative of the relaxation mechanisms \cite{GSA}.  \\
\indent Indeed, our TDII measurements of the fluorescent emission from the 
cold atoms  exhibit such  oscillations (Fig.\ref{g2}(b)) --   for large 
detunings, five full oscillations are seen while for small detunings 
barely one or two are. 
That the oscillations are reduced in prominence as detuning decreases may be  understood in terms of the 
temperature of the atoms.  Small detunings
of the cooling beams result in a hotter collection of atoms,  and therefore result in 
increased  inter-atomic collisions that decohere the system rapidly.  It is thus evident that  
TDII measurements can help determine the relative strengths of various 
relaxation mechanisms as functions of different physical parameters. \\
\indent For the collection of cold atoms in our experiments, signatures of the coherent 
processes in the \gt 
die out within  delays of $\sim150$ns, and spontaneous emission 
is expected to  have caused atoms to make transitions to the ground 
state within a few lifetimes of the excited state.  For
delay times larger than this, \gt  
is dominated by effects due to the  scattering by moving atoms.  While the 
collection of atoms under study is cooled to  $\sim 100 \mu$K, where the 
Doppler width reduces below the natural linewidth, it may seem surprising that 
the effect of the velocity distribution is seen in the scattering. Once again, the
power of the second order correlation becomes evident, as \gt may be 
interpreted as 
the measure of the probability  of detecting a second  photon scattered by an  
atom with velocity $v$, within a time $\tau$  of having  detected one such 
photon. When the velocity spread of atoms is large, as at higher temperatures, 
the probability for such an event is low, and thus \gt will fall more rapidly 
towards unity compared to  the case for lower temperatures.  Thus, \gt at large 
time delays (in this case delays larger than $\sim$500ns)  may be used to determine 
the temperature of the ensemble. 
The elastically scattered light has a Doppler profile determined by the velocity 
distribution of atoms\cite{Westbrook, Bali}. In the six-beam configuration, denoting by  $\alpha_j ( = 2(1-cos
\theta_j)$ ) the dependence of the Doppler spread of the $j^{th}$ beam  on its scattering angle 
$\theta_j$, and by $A_j$ the weight factor for the $j^{th}$ beam appropriate for  its
intensity and its  polarisation and angle dependent elastic scattering cross-section \cite{Westbrook, Bali},
\begin{equation}
g^{(1)}_D(\tau) =   \sum_{j=1}^{6} A_i exp(-\alpha_j \frac{\omega_o^2 k_B T \tau^2}{2mc^2})
\label{Doppler}
\end{equation}
 Here $g^{(1)}_D(\tau)$ represents the Doppler contribution to the first order 
correlation function and $\omega_o$, c,  $k_B$, T and m  are the frequency and speed of 
light, the  
Boltzman constant,   the temperature of the ensemble and the  mass of the atom,
respectively. Using this in conjunction with the relation  \footnote {For finite (non-zero)  size of source and detector, a factor S is introduced in Eq. \ref{eq-g2-g1} (see, for example, \cite{Nakayama} )}
\begin{eqnarray}
g^{(2)}(\tau)&= 1 + S \,\, |  g^{(1)}(\tau)  |  ^2 \\
&\,\,\,\approx 1 + S \,\, |  g^{(1)}_D(\tau)  |  ^2  \,\,\, \text{for large $\tau$}
\label{eq-g2-g1-D}
%\end{align}
\end{eqnarray}
where S depends on the spatial  coherence of the light detection system, the
temperature T may be estimated from the experimentally obtained time-delayed 
intensity correlation function.  
Fig. \ref{temp} displays the experimentally 
obtained valued for \gt as function of $\tau$ for different detunings $\delta$
of the cooling beam. The temperature of the ensemble is 
determined for each detuning of the cooling beam  by fitting  the experimental data with the corresponding 
\gt  curve obtained using Eq. \ref{Doppler} and  Eq.\ref{eq-g2-g1-D}.  
As seen in the figure, the data for the different detunings fit quite well with the 
respective curves. It may be noted that  the same parameters ($S, \alpha_i, A_i$) are used for all curves. 
The values of temperature thus obtained ($T_{TDII}$), on comparison 
with  
the  temperature obtained by the trap oscillation method ($T_{TO}$)
show fairly good agreement (Fig.\ref{temp-compare}),  
with $T_{TDII}$ being slightly lower in all cases. 

 We now turn our attention to the remaining observation - the increase in \go 
with detuning. We attribute this to the timing resolution in our experiment. The larger the detuning, 
the colder the collection of atoms, and hence the slower the decay in coherence. 
Thus the timing resolution of 5ns appears adequate for large detunings. For hotter 
atoms obtained at lower detunings, the averaging effect of the time bin becomes discernible, as it is now a larger fraction of the coherence time. 

As pointed out earlier, though light form a thermal collection of atoms is 
expected to exhibit a second-order intensity correlation of 2, this value could not 
be experimentally obtained  till very recently; in fact, the present work and that of Nakayama and 
coworkers \cite{Nakayama} are the only two reports of this.  Further, there has 
been skepticism on being able to obtain a good measure of bunching, and of 
being able to see coherent effects in TDII measurements from a collection of large number of atoms \cite{Gomer}.  Several 
factors have 
contributed to the difficulty in observing the theoretically predicted behaviour. 
All researchers have stressed on the need for good timing 
resolution, which in the present case is 5ns. However, small time bins 
necessitate longer acquistion times to obtain good statistics, making the 
experiment long and tedious. Light has to be collected over a single spatial 
coherence region, contributing  to further  reduction in photon counts. 
These factors require the experimental setup  to be extremely stable, 
and the conditions repeatable over nearly ten hours. Another factor known to 
degrade the observation of bunching is the presence of a magnetic field  
\cite{Bali}, because of which 
 all measurements hitherto had atoms
cooled in a MOT and then transferred to a molasses, either by switching off the magnetic field (\cite{Bali, Aspect}), or by transporting the atoms to another 
vacuum chamber \cite{Nakayama}. Our experiment, however, has been carried 
out in-situ, with atoms in a MOT, with all cooling and  repumper beams and the 
quadrupolar magnetic field present.  Good mechanical isolation of the setup and 
temperature stability of 
the environment ensured that the lasers remained locked for the entire duration 
of the experiment. Constant monitoring of the MOT fluorescence allowed for 
corrective measures, which, however, were not required. The 
diffraction-limited 
collection lens placed within the MOT chamber, in close proximity to the cold atoms, 
and the subsequent spatial filtering enabled us to collect light from a small
region of the MOT, over which the magnetic field was  uniform 
within 2mG, eliminating broadening due to Zeeman shift. Likewise, the low 
temperature, and the thus the sub-natural Doppler width ensured that the Rabi 
frequency is the same for all atoms.  Further, the small size of the cold cloud 
($400\mu$m across), 
the low number of atoms ensured that reabsorption of the emitted light 
was negligible.  This allowed us to detect coherent effects like  
Rabi oscillations. In  an earlier study, single atom dynamics was
probed by photon-photon 
correlation in an optical dipole trap\cite{Gomer}, where one, two, or three 
atoms were held trapped.  In the present experiment, the number of atoms contributing to the collected light is three orders of 
magnitude higher. Further, 
atoms move in and out of 
the region from which light is collected, due to the thermal motion, and the 
superimposed trap oscillation in the quadrupolar magnetic field. The transit 
time of an atom (in the absence of a collision that expels it from this region), 
is estimated to be $\sim 10\mu$s.  The power of TDII 
is brought out in the present study, where,   
despite the sample  being a thermal collection of several thousand atoms,  coherent 
dynamics are revealed.\\
%\section {summary}
\indent In conclusion, we have perfomed Time-Delayed Intensity Interferometry 
with light emitted by an ultracold atomic ensemble in a steady state MOT. The 
collection of cold atoms 
is a source of bunched light, where bunching is introduced by spontaneous 
emission. Well defined, but decaying Rabi oscillatons were seen at small time 
delays ($< 150$ns) that give way  to an exponential decay at larger time delays. 
It is thus seen that  TDII measurements enable the study of coherent and incoherent 
dynamics of the system, providing  a  relative measure of the 
various dynamical processes occuring at different time scales, even from a 
thermal ensemble of a large number of independent atoms. \\
Acknowledgements : We thank Ms. M.S. Meena for her
efficient help in electronics related to the setting up of the
MOT. We gratefully acknowledge Prof. G. S. Agarwal for several 
detailed discussions and for the theoretical treatment presented here.\\


\begin{thebibliography}{00}

\bibitem{HBT} R. Hanbury Brown and R. Q. Twiss, Nature {\bf 177}, 27 (1956).

\bibitem{Baym} Gordon Baym, Acta Physica Polonica, {\bf29}, 1839  (1998). 


\bibitem{particle-physics}R. Lednicky, Brazilian Journal of Physics,  {\bf 37}, 939, 2007

\bibitem{stellar}C. Foellmi, Astr. Astrophy., {\bf 507}, 1719-1727 (2009). 
 

\bibitem{Bayer}M. Assmann, F. Veit, J. Tempel,T. Berstermann,
H. Stolz, M. van der Poel, J. M. Hvam, and M. Bayer, Opt. Exp., {\bf 18},   20229 (2010).

\bibitem{DLS}http, www.lsinstruments.ch,technology,dynamic,light,
scattering,dls

\bibitem{Hg1} D.T. Philips, Herbert Kleiman and Sumner P. Davis, Phys. Rev. {\bf153}, 113 (1966).

\bibitem{GroundGlass} W. Martienssen and E. Spiller, Am J Phys (1964). 

\bibitem{Anders}A. Martin, O. Alibert, J.C. Flesch, J. Samuel, Supurna Sinha, S. Tanzilli, and A. Kastberg,  EPL {\bf 1}, 10003 (2012).

\bibitem{AB-EPL}Nandan Satapathy, Deepak Pandey, Poonam Mehta, Supurna Sinha, Joseph Samuel, and Hema Ramachandran,  EPL {\bf 97}, 50011 (2012).

\bibitem{JOSA} Nandan Satapathy, Deepak Pandey, Sourish Banerjee, and Hema Ramachandran, J. Opt. Soc. Am. A, {\bf 30}, 910 (2013). 

\bibitem{EPJP}Deepak Pandey, Nandan Satapathy, Buti Suryabrahmam, J. Solomon Ivan and Hema Ramachandran, Eur. Phys. J. Plus {\bf129}, 115 (2014). 

\bibitem{Aspect}C. Jurczak, B. Desruelle, K. Sengstock, J.-Y. Courtois, C. I. Westbrook, and A. Aspect,  Phys. Rev. Lett., {\bf77}, 1727 (1996)

\bibitem{Bali}S. Bali, D. Hoffmann, J. Siman, and T. Walker, Phys.  Rev. A {\bf53},  3469 (1996).  

\bibitem{Nakayama}K. Nakayama, Y.  Yoshikawa, H. Matsumoto,
Y. Torii1, and T. Kug,  Opt. Exp. {\bf18}, 6604 (2010).

\bibitem{arXiv} Manuscript under preparation. 

%\bibitem{Kaiser}C. Jurczak, K. Sengstock, R. Kaiser, N. Vansteenkiste, C.J. Westbrook, A.Aspect, Optics Comunications, 115 (1995), 480-484. 


%\bibitem{Jessen}

\bibitem{trap-osc}P. Kohns, P. Buch, W. Suptitz, C. Csambal and W. Ertmer, 
Europhys. Lett., {\bf 22}, 517 (1993). 

\bibitem{GSA} Chapter 13, Quantum Optics by Girish S. Agarwal, Cambridge University Press, 2012. 


\bibitem{Gomer}V. Gomer, F. Strauch, B. Ueberholz, S. Knappe, and D. Meschede, Phys. Rev. A {\bf 58}, R1657 (1998). 

%\bibitem{Mollow} B.R. Mollow, Phys. Rev. {\bf 188}, 1969 (1969). 


\bibitem{Westbrook} C. I. Westbrook, R. N. Watts, C. E. Tanner, S. L. Rolston, W. D. Phillips, and P. D. Lett, Phys. Rev. Lett., {\bf 65}, 33 (1990). 

\end{thebibliography}
\end{document}